\documentclass[twocolumn,longauthor]{aastex61}

\usepackage{aas_macros}
\usepackage[lined]{algorithm2e}
\usepackage{lineno}
\usepackage{verbatim, amsmath, amsfonts, amssymb,amssymb}
\usepackage{pifont}
\usepackage{enumitem}
\usepackage{booktabs}
\usepackage{multirow}
\usepackage[utf8]{inputenc}
\usepackage[T1]{fontenc}
\usepackage{empheq}
\usepackage[skins,theorems]{tcolorbox}
\usepackage{latexsym}
\usepackage{bigints}

\tcbset{highlight math style={enhanced,
 colframe=cyan!10!gray,colback=gray!5!white,arc=4pt,boxrule=1pt,
  drop fuzzy shadow}}

  {\empheq[box=\tcbhighmath]{align}}
  {\endempheq}
  
\definecolor{red2}{HTML}{7f0000}

\received{\today}
\submitjournal{ApJ}

\shorttitle{Thermonuclear reaction rates and primordial nucleosynthesis}
\shortauthors{Iliadis and Coc}

\begin{document}

\title{Thermonuclear reaction rates and primordial nucleosynthesis}

\correspondingauthor{Christian Iliadis}
\email{iliadis@unc.edu}

\author[0000-0003-2381-0412]{Christian Iliadis}
\affiliation{Department of Physics \& Astronomy, University of North Carolina at Chapel Hill, NC 27599-3255, USA}
\affiliation{Triangle Universities Nuclear Laboratory (TUNL), Durham, North Carolina 27708, USA}

\author{Alain Coc}
\affiliation{CNRS/IN2P3, IJCLab, Universit\'e Paris-Saclay, B\^atiment, 104, F-91405 Orsay Campus, France}

\begin{abstract}
Assuming the best numerical value for the cosmic baryonic density and the existence of three neutrino flavors, standard big bang nucleosynthesis is a parameter-free model. It is important to assess if the observed primordial abundances can be reproduced by simulations. Numerous studies have shown that the simulations overpredict the primordial $^7$Li abundance by a factor of $\approx$ $3$ compared to the observations. The discrepancy may be caused by unknown systematics in $^7$Li observations, poorly understood depletion of lithium in stars, errors in thermonuclear rates that take part in the lithium and beryllium synthesis, or physics beyond the standard model. Here, we focus on the likelihood of a nuclear physics solution. The status of the key nuclear reaction rates is summarized. Big bang nucleosynthesis simulations are performed with the most recent reaction rates and the uncertainties of the predicted abundances are established using a Monte Carlo technique. Correlations between abundances and reaction rates are investigated based on the metric of mutual information. The rates of four reactions impact the primordial $^7$Li abundance: $^3$He($\alpha$,$\gamma$)$^7$Be, d(p,$\gamma$)$^3$He, $^7$Be(d,p)2$\alpha$, and $^7$Be(n,p)$^7$Li. We employ a genetic algorithm to search for simultaneous rate changes in these four reactions that may account for all observed primordial abundances. When the search is performed for reaction rate ranges that are much wider than recently reported uncertainties, no acceptable solutions are found. Based on the currently available evidence, we conclude that it is highly unlikely for the cosmological lithium problem to have a nuclear physics solution.
\end{abstract}




\section{Introduction} 
\label{sec:intro}
The standard model of the big bang rests on the three observational pillars of big bang nucleosynthesis \citep{Gamow1948,cyburt16,Mat17}, cosmic expansion \citep{Riess1998,Peebles2003}, and cosmic microwave background radiation \citep{Spergel2007,Planck2016}. Primordial nucleosynthesis took place during the first $20$ minutes after the big bang and probes the cosmological model at early time. Nucleosynthesis in the early universe proceeded at temperatures and densities below $1$~GK and 10$^{-5}$~g/cm$^3$, respectively. It produced the nuclides $^1$H (or H), $^2$H (or D), $^3$He, $^4$He, and $^7$Li, which represent the building blocks for the subsequent evolution of matter in the universe.

Primordial abundances have been measured in the proper astrophysical site with high precision in recent years. The cosmological deuterium abundance is determined by observing damped Ly-$\alpha$ systems at high redshift, resulting in a primordial number abundance ratio relative to hydrogen of (D/H)$_p^{obs}$ $=$ $(2.527\pm0.030)\times 10^{-5}$ \citep{Cooke2018}. The nuclide $^3$He has not been observed outside of the Galaxy since its abundance is very small. Also, the galactic chemical evolution of $^3$He is uncertain because it is both produced and destroyed inside stars. Therefore, only an upper limit has been established from observations, ($^3$He/H)$_p^{obs}$ $\leq$ $(1.1\pm0.2)\times 10^{-5}$ \citep{Bania:2002wn}. The primordial $^4$He abundance is deduced from measurements of metal-poor extragalactic H~II regions, resulting in a nucleon fraction of Y$_p^{obs}$ $=$ $0.2449\pm0.0040$ \citep{Aver2015}. The cosmological $^7$Li abundance is estimated from observations of low-metallicity stars in the halo of the Galaxy, where the lithium abundance is nearly independent of metallicity. The result is ($^7$Li/H)$_p^{obs}$ $=$ $(1.58\pm0.31)\times 10^{-10}$ \citep{sbordone10}. The uncertainties for the measured primordial $^2$H, $^4$He, and $^7$Li abundances correspond to values of 1.2\%, 1.6\%, and 19.6\%, respectively. These values are uncorrelated since each of these nuclides is observed in different astronomical objects by independent measurement techniques. The observations will be most useful for constraining cosmology if predicted primordial abundance uncertainties can be reduced to a level similar to that of the experimental ones. 

The precision in the cosmological parameters entering the model of big bang nucleosynthesis has also been improved significantly in recent years. In particular, the measurement of anisotropies in the cosmic microwave background radiation has determined the cosmic baryonic density, $\Omega_b \cdot h^2$ $=$ $0.02233\pm0.00015$ \citep{collaboration2018planck}, with a precision of less than $1$\%. If we assume, in addition, $N_\nu$ $=$ $3$ for the number of neutrino flavors, the standard model of big bang nucleosynthesis becomes a parameter-free theory. These results establish the expansion rate of the early universe and the prevailing thermodynamic conditions (i.e., temperature and density). All that is required to simulate the primordial abundances is to numerically solve the network of coupled differential equations describing the abundance evolution of the light nuclides. The numerical results depend sensitively on the magnitudes and associated uncertainties of the thermonuclear reaction and weak interaction rates (see, e.g., \citet{Ser04}). 

Many studies \citep{Sch98,Steigman:2007ky,Oli10,PhysRevD.98.030001} have found that the predicted primordial abundances, with one exception, conform broadly with the experimental values. This result is significant because the abundances span nine orders of magnitude. Consequently, big bang nucleosynthesis is a sensitive tool for probing the physics of the early universe as well as for constraining physics beyond the standard model \citep{Ioc09}. The exception is the primordial $^7$Li abundance for which the simulated value exceeds the observational result by a factor of about three. This is the {\it cosmological lithium problem} and it has not found a satisfactory solution yet. The discrepancy may be caused by unknown systematics in the $^7$Li observations, poorly understood depletion of lithium in stars, errors in thermonuclear rates of reactions that take part in the lithium and beryllium synthesis, or physics beyond the standard model. Solutions involving exotic physics almost always lead to an overproduction of deuterium incompatible with observations \citep{Olive12,Kus14,Coc:2015gt}. Solutions involving stellar physics require a uniform {\em in situ} lithium depletion \citep{Mic84,Korn06} over wide metallicity
and efffective surface temperature ranges \citep{Agu19,Spi15}. See \citet{Fie11,cyburt16,Fields_2020,Davids20} for reviews.

The goal of this work is to investigate the nuclear physics aspects of this persistent problem. In Section~\ref{sec:nuclear}, we summarize the most recent information regarding the thermonuclear rates of key reactions. Our reaction network, thermodynamic conditions, and initial abundances are discussed in Section~\ref{sec:network}. Results from a network Monte Carlo procedure are presented in Section~\ref{sec:mc}. In Section~\ref{sec:correlations}, we study correlations between reaction rates and simulated abundances. A genetic algorithm is employed in Section~\ref{sec:bias} to investigate whether simultaneous rate changes of key nuclear reactions can account for the observed primordial abundances. A concluding summary is given in Section~\ref{sec:summary}. In the Appendix, we provide information on the rates of selected nuclear reactions. 

\section{Nuclear processes: Bayesian reaction rates} 
\label{sec:nuclear}
The twelve most relevant nuclear processes taking part in primordial nucleosynthesis are listed in column 1 of Table~\ref{tab:rates}, together with their rate uncertainty at $1$~GK (column 2), the reference for the rates adopted here (column 3), and comments on the data treatment (column 4). 

\begin{deluxetable*}{lcll}
\tablecaption{Main nuclear processes in primordial nucleosynthesis.\tablenotemark{a} \label{tab:rates}}
\tablewidth{2\columnwidth}
\tablehead{ Process  &   Uncertainty (\%)\tablenotemark{b}  & Most recent reference & Comments } \startdata
   n $\leftrightarrow$ p            &      0.09  &      \citet{Pitrou2018}              & theory and neutron lifetime \\ 
   p(n,$\gamma$)d                   &      0.4   &      \citet{Ando2006}                & MCMC analysis of data  \\ 
   d(p,$\gamma$)$^3$He              &      3.7   &      \citet{iliadis16}               & Bayesian data analysis  \\ 
   d(d,p)t                          &      1.1   &      \citet{gomez17}                 & Bayesian data analysis  \\ 
   d(d,n)$^3$He                     &      1.1   &      \citet{gomez17}                 & Bayesian data analysis  \\ 
   $^3$He(n,p)t                     &      1.6   &      \citet{Descouvemont2004}        & $\chi^2$ data fitting  \\ 
   $^3$He(d,p)$\alpha$              &      1.2   &      \citet{deSouza:2019gi}          & Bayesian data analysis  \\ 
   t(d,n)$\alpha$                   &      0.3   &      \citet{deSouza:2019gf}          & Bayesian data analysis  \\ 
   $^3$He($\alpha$,$\gamma$)$^7$Be\tablenotemark{c}  &      2.4   &      \citet{iliadis16}               & Bayesian data analysis  \\ 
   t($\alpha$,$\gamma$)$^7$Li       &      4.3   &      \citet{Descouvemont2004}        & $\chi^2$ data fitting \\ 
   $^7$Be(n,p)$^7$Li                &      2.1   &      \citet{de_Souza_2020}           & Bayesian data analysis  \\ 
   $^7$Li(p,$\alpha$)$\alpha$       &      2.5   &      \citet{Descouvemont2004}        & $\chi^2$ data fitting  \\ 
\enddata
\tablenotetext{a}{The symbols p, d, t, and $\alpha$ denote the nuclides $^1$H, $^2$H, $^3$H, and $^4$He, respectively.}
\tablenotetext{b}{Rate uncertainty at $1$~GK.}
\tablenotetext{c}{See Section~\ref{sec:3heag}.}
\end{deluxetable*}

The rates of the weak interactions that convert neutrons to protons, and vice versa, $n$ $\leftrightarrow$ $p$, have recently been discussed and computed by \citet{Pitrou2018}, who took into account the effects of radiative corrections (including the effects of finite nucleon mass, finite temperature, weak magnetism, quantum electrodynamics, and incomplete neutrino decoupling). The weak rates scale inversely with the free-neutron lifetime\footnote{This value is based on the results of seven experiments compiled by the Particle Data Group \citep{PhysRevD.98.030001} and three more recent measurements \citep{PhysRevC.97.055503,Pattie627,Ezhov:2018vx}. A slightly different value of $\tau$ $=$ $879.5\pm0.8$~s was used in \citet{Pitrou2018}.}, for which we adopt a value of $\tau$ $=$ $879.7\pm0.8$~s, and, thus, have an uncertainty of $\leq$ $0.1$\%. 

The p(n,$\gamma$)d reaction rate has been calculated precisely using effective field theory \citep{Ando2006}. The model parameters were constrained, using a Markov chain Monte Carlo procedure, by cross section data for neutron capture, deuteron photodissociation, and photon analyzing powers. The reported rate uncertainty is $0.4$\%.

Ten reactions listed in Table~\ref{tab:rates} have been measured directly in the laboratory at the energies of astrophysical interest. The rates for the $^3$He(n,p)t, t($\alpha$,$\gamma$)$^7$Li, and $^7$Li(p,$\alpha$)$\alpha$ reactions are adopted from \citet{Descouvemont2004}. These were obtained using frequentist statistics and $\chi^2$ optimization, with the implicit assumption of normal likelihoods and approximate treatments of systematic uncertainties. The reported rate uncertainties for these three reactions amount to 1.6\%, 4.3\%, and 2.5\%, respectively. These  are small enough that they do not impact big bang nucleosynthesis predictions.

The cross section data for the remaining seven reactions have been analyzed recently using Bayesian hierarchical models. For d(p,$\gamma$)$^3$He, d(d,n)$^3$He, d(d,p)t, and $^3$He($\alpha$,$\gamma$)$^7$Be, microscopic theories of nuclear reactions provided the  cross section energy dependence, while the absolute cross section normalization was found from fitting the data within the Bayesian framework \citep{iliadis16,gomez17}. For $^3$He(d,p)$\alpha$, t(d,n)$\alpha$, and $^7$Be(n,p)$^7$Li, R-matrix theory was implemented into the Bayesian model to fit the data \citep{deSouza:2019gi,deSouza:2019gf,de_Souza_2020}. The Bayesian method provides statistically rigorous rate probability densities that can be used in Monte Carlo reaction network studies to derive meaningful simulated primordial abundance uncertainties. 


\citet{Fields_2020} prefer to {\it ``use smooth polynomial fits to the data...by construction this ensures that the fits match the data as well as possible given the uncertainties.''} In the present work, we instead prefer to use for our adopted rates (Table~\ref{tab:rates}) well-established nuclear theory (microscopic models or R-matrix theory) as a foundation for the data analysis. We are not implying that one method is superior to the other, only that there is a range of reasonable options. We also prefer to adopt rates obtained using Bayesian techniques because they allow for a rigorous inclusion of statistical and systematic sources of uncertainties in the data analysis. For a recent example of such a study, see the analysis of $^7$Be(n,p)$^7$Li data by \citet{de_Souza_2020}.

A few other reaction rates that impact primordial nucleosynthesis, e.g., for d(n,$\gamma$)t, $^3$He(t,d)$\alpha$, and $^7$Be(d,p)2$\alpha$, have been adopted from the literature. They are listed in Table~\ref{tab:rates2} and will be discussed in the text below and in the Appendix. The rates of all other reactions in our network (see Section~\ref{sec:network}) have been adopted from statistical nuclear model calculations. See \citet{Sallaska2013} for details. We assigned a factor of $10$ uncertainty to the latter rates.

\begin{deluxetable}{lcll} 
\tablecaption{Other nuclear reactions in primordial nucleosynthesis considered in the present work.\tablenotemark{a}}
\label{tab:rates2}
\tablewidth{\linewidth}
\tablehead{
Process  &   \multicolumn{2}{c}{Uncertainty\tablenotemark{b}}  & Most recent reference \\
         &   Reported\tablenotemark{c}     &    Adopted\tablenotemark{d}          &} 
\startdata
   d(n,$\gamma$)t\tablenotemark{e}          &      8.0\%    &  factor 2     &    \citet{Nagai:2006kz}   \\ 
   $^3$He(t,d)$\alpha$\tablenotemark{f}     &      30\%     &  factor 10    &    \citet{CF88}           \\ 
  $^7$Be(d,p)2$\alpha$\tablenotemark{g}     &     factor 3  &  factor 3     &    \citet{Rijal:2019bh}   \\
\enddata
\tablenotetext{a}{The symbols p, d, t, and $\alpha$ denote the nuclides $^1$H, $^2$H, $^3$H, and $^4$He, respectively.}
\tablenotetext{b}{Rate uncertainty at $1$~GK.}
\tablenotetext{c}{Reported in original work.}
\tablenotetext{d}{Adopted here.}
\tablenotetext{e}{See Section~\ref{sec:dng}.}
\tablenotetext{f}{See Section~\ref{sec:he3td}.}
\tablenotetext{g}{See Section~\ref{sec:be7dp}.}
\end{deluxetable}

\section{Reaction network, thermodynamic conditions, and initial abundances}
\label{sec:network}
We compute primordial nucleosynthesis using a reaction network consisting of $50$ nuclides, ranging from p, n, $^4$He, to $^{22}$F. These are linked by $466$ nuclear processes, such as particle captures, weak interactions, light-particle reactions, etc. Thermonuclear reaction rates are adopted  from STARLIB v6.8 (04/2020).\footnote{The STARLIB site has moved to \url{https://starlib.github.io/Rate-Library/}.} The library is in tabular format and lists reaction rates and rate factor uncertainties on a grid of temperatures between $1$~MK and $10$~GK \citep{Sallaska2013}. Except for a few recently updated reaction rates (see text), the network is similar to the one used by \citet{Pitrou2018}.

\begin{figure}
\includegraphics[width=1\linewidth]{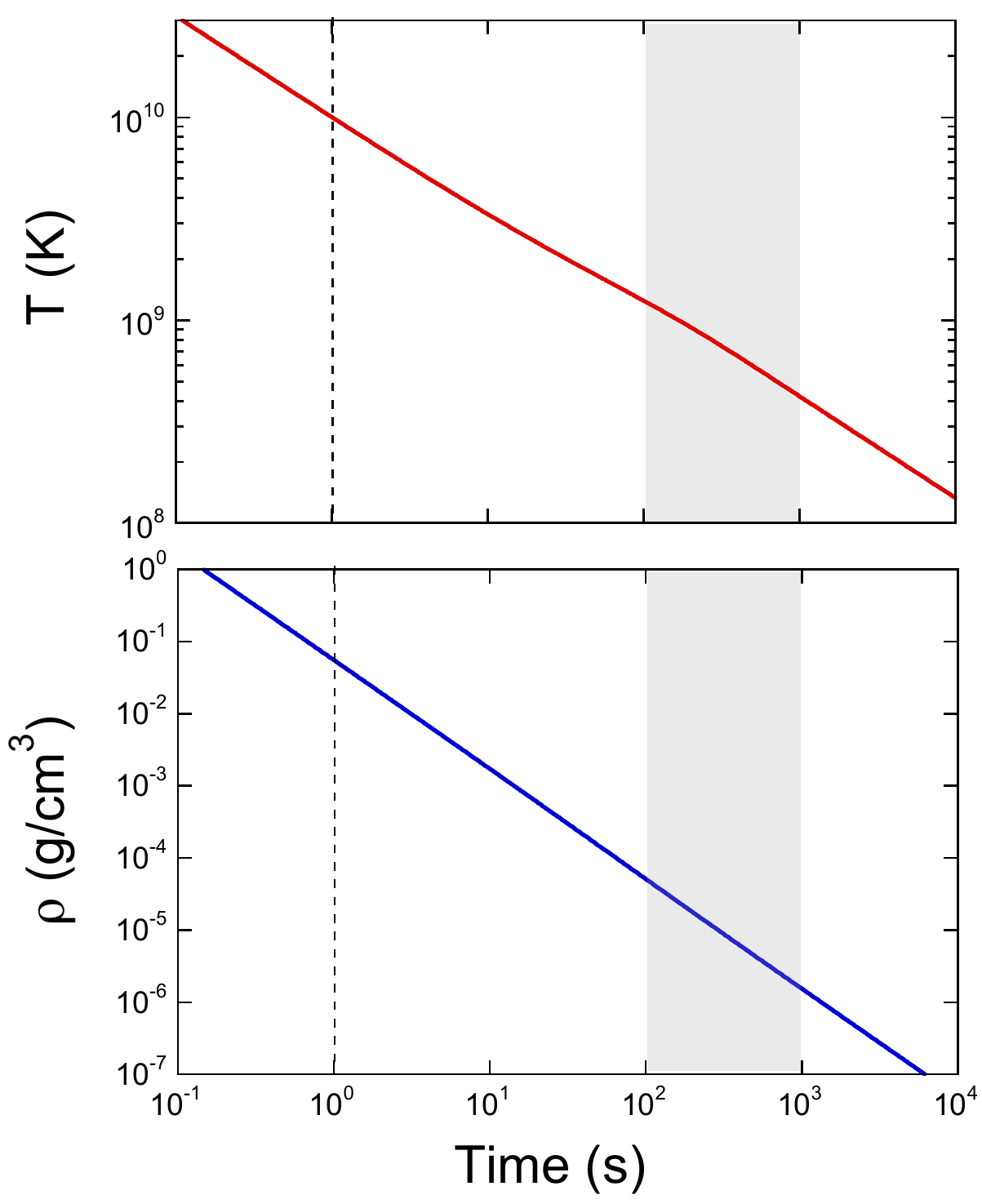}
\caption{Temperature (top) and density (bottom) evolution of the early universe. We start each reaction network calculation at time $t$ $=$ $1$~s after the big bang (vertical dashed lines). The most significant primoridal nucleosynthesis occurs between $100$~s and $1000$~s (shaded regions). Results are adopted from \citet{Pitrou2018}.}
\label{fig:bbnCond}
\end{figure}

Assuming that the baryonic density is known (Section~\ref{sec:intro}), the thermodynamic conditions (temperature and density) in the early universe are determined by the expansion rate. The evolutions of temperature and density, adopted from \citet{Pitrou2018}, are presented in Figure~\ref{fig:bbnCond}. We start the network calculations at a time of $1.0$~s after the big bang, indicated by the vertical dashed lines in both panels. At that time, the temperature and density amount to $T$ $=$ $9.93$~GK and $\rho$ $=$ $0.0537$~g/cm$^3$, respectively. We end each calculation at a time of $10^5$~s after the big bang. 

\begin{figure}
\includegraphics[width=1\linewidth]{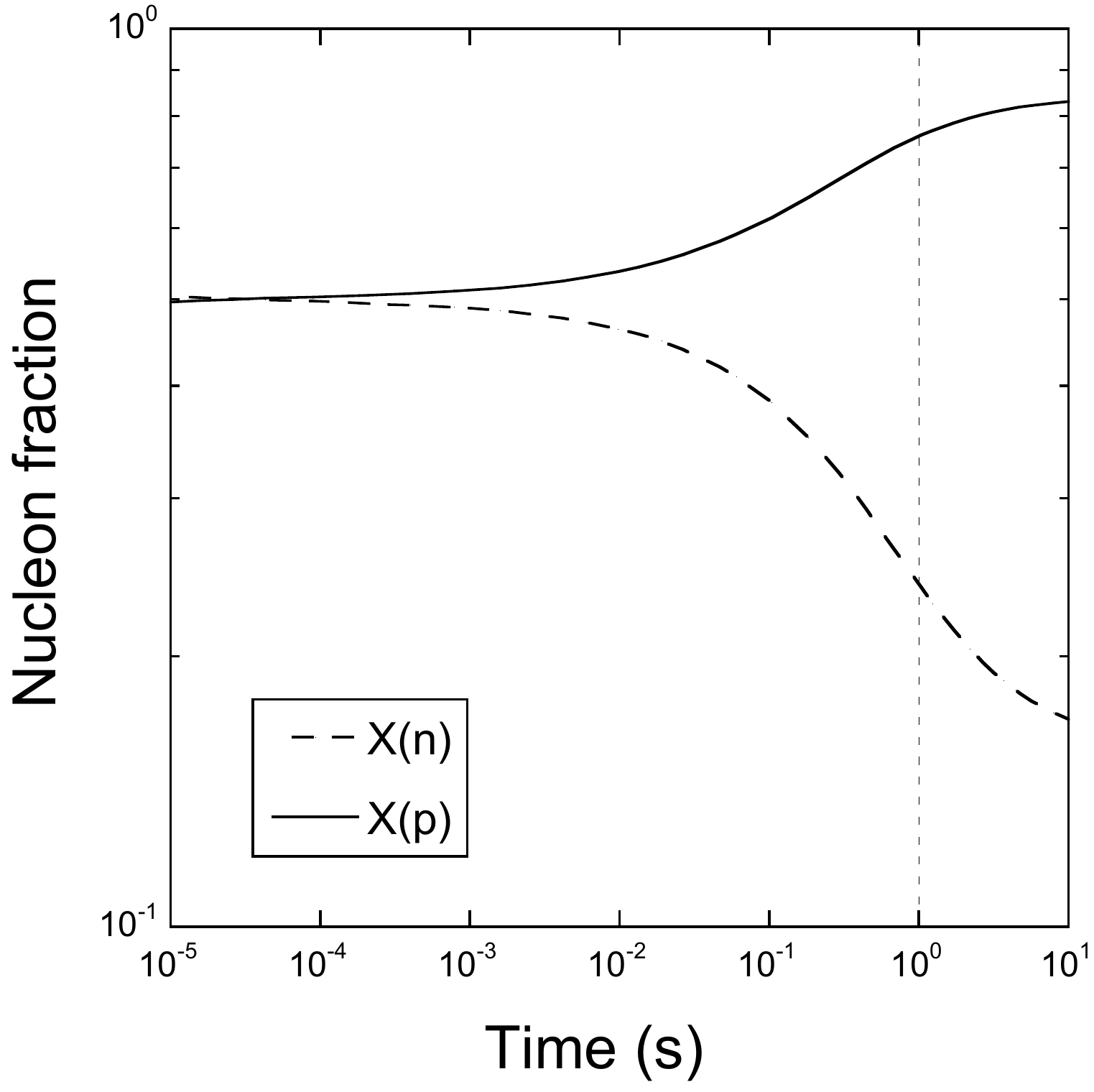}
\caption{Nucleon fractions of protons (solid line) and neutrons (long-dashed line) for the first $10$~s after the big bang. We start each reaction network calculation at time $t$ $=$ $1$~s after the big bang (vertical dashed line). Results are adopted from \citet{Pitrou2018}.}
\label{fig:bbnX}
\end{figure}

The last ingredients needed for setting up the reaction network are the neutron and proton nucleon fractions at the beginning of the calculation. For times $t$ $\leq$ $0.1$~s, with temperatures of $T$ $\geq$ $30$~GK, all particles (photons, electrons, positrons, protons, neutrons, neutrinos, and antineutrinos) are in thermal equilibrium and the neutron-to-proton number abundance ratio is given by the Boltzmann distribution, $N_n/N_p$ $=$ $\exp{(-Q/kT)}$, where $Q$ $=$ $1.29333205(48)$~MeV is the energy equivalent of the neutron-proton mass difference \citep{RevModPhys.88.035009} and $k$ is the Boltzmann constant. As the universe expands and cools, neutrons and protons fall out of equilibrium because the weak rates ($n$ $+$ $\nu$ $\leftrightarrow$ $p$ $+$ $e^-$, $n$ $\leftrightarrow$ $p$ $+$ $e^-$ $+$ $\overline{\nu}$, $n$ $+$ $e^+$ $\leftrightarrow$ $p$ $+$ $\overline{\nu}$) 
become comparable to, and then slower than, the expansion rate. Consequently, the neutron and proton abundances freeze out and their ratio continues to change as a result of free neutron decay. Figure~\ref{fig:bbnX} displays the nucleon fractions of protons and neutrons as solid and long-dashed lines, respectively. The vertical dashed line indicates the time at which we start the network calculations, when the neutron and proton nucleon fractions are $X_n$ $=$ $0.23948$ and $X_p$ $=$ $0.76052$, respectively.

Primordial nucleosynthesis starts about $100$~s after the  big bang, when the temperature has declined to about $1$~GK. This temperature is sufficiently low for the first nuclear reaction, p(n,$\gamma$)d, to become faster than its reverse photodisintegration, which until this point prevented the production of heavier nuclei. The nuclear transformations result mainly in the synthesis of $^4$He, because it has the highest binding energy among all hydrogen and helium nuclides. It becomes the most abundant species (after $^1$H), while all other nuclides are produced with much smaller abundances. The nucleosynthesis continues until about $1000$~s after the big bang (shaded regions in Figure~\ref{fig:bbnCond}). For later times, the temperature and density in the early universe were too small for thermonuclear reactions to change primordial abundances. 

\section{Simulated abundances: Monte Carlo studies} 
\label{sec:mc}
We will next investigate the answer to the following question: {\it Given a set of recommended reaction rates and their uncertainties, and assuming that these were obtained by taking into account all known statistical and systematic effects in the measured data, what are the best estimates of the simulated primordial abundances?} This question can best be answered using a Monte Carlo network technique \citep{Oli00, Nol00,Coc02}. The following steps are involved. First, the rates of all $466$ reactions in the network are randomly sampled (see below). Second, the network is numerically solved using this set of sampled rates. Third, the first two steps are repeated $n$ times to collect an ensemble of final abundance yields; $n$ must be sufficiently large for statistical fluctuations to become much smaller than the widths of the obtained abundance distributions. Fourth, final primordial abundances are extracted from the accumulated abundance probability densities. 

The nuclear interaction rates are sampled using lognormal rate probability densities \citep{Coc_2014,Iliadis:2015gp}. For a given reaction, $j$, and temperature, $T$, the rate probability density is expressed as
\begin{equation}
f[x(T)_j] = \frac{1}{\sigma \sqrt{2\pi}} \frac{1}{x(T)_j} e^{-[\ln x(T)_j - \mu(T)_j]^2/[2\sigma(T)_j^2]}
\end{equation}
where the lognormal parameters $\mu$ and $\sigma$ determine the location and the width of the distribution, respectively. For a lognormal probability density, rate samples, $x_i$, are drawn using \citep{Longland:2012kv} 
\begin{equation}
\label{eq:ratesample}
x(T)_{ij} = x(T)_{med,j} [f.u.(T)]_j^{p(T)_{ij}}
\end{equation}
where $x_{med}$ and $f.u.$ are the median rate value and the rate factor uncertainty, respectively. Both of these are listed in columns 2 and 3 of STARLIB, respectively. The quantity $p_{ij}$ is a normally distributed random variable, i.e., it is given by a Gaussian distribution with an expectation value of zero and a standard deviation of unity. Note that $f.u.(T)^{p(T)}$, and not $p(T)$, is the factor by which the sampled reaction rate is modified compared to its median value. We will refer to $p(T)$ as the {\it rate variation factor}.

We sample this rate variation factor for a given network run and nuclear reaction only once, i.e., $p(T)_{ij}$ $=$ $p_{ij}$ has the same value at all temperatures. This assumption was found to reproduce the abundance uncertainties arising from more complex sampling schemes \citep{Longland:2012kv}. It implies that the rate samples still depend on temperature, since Equation~(\ref{eq:ratesample}) takes the temperature dependence of the factor uncertainty, $f.u.(T)$, into account. Also, the rates of corresponding forward and reverse reactions are not  sampled independently, because they are subject to the same value of the rate variation factor in a given network calculation.

The Monte Carlo procedure has major advantages compared to varying rates one-by-one in sequential network runs. It is straightforward to estimate primordial abundances by adopting the 16th, 50th, and 84th percentiles of the resulting abundance probability densities. The impact of a given reaction rate uncertainty on the nucleosynthesis can be quantified by storing the values of $p_{ij}$ for each sample reaction network run. A scatter plot for the final abundance of a given nuclide versus the sampled value of $p_{ij}$ can then be inspected for correlations. Below, we will present results obtained by computing 10,000 Monte Carlo reaction network samples.

\begin{figure}
\includegraphics[width=0.9\linewidth]{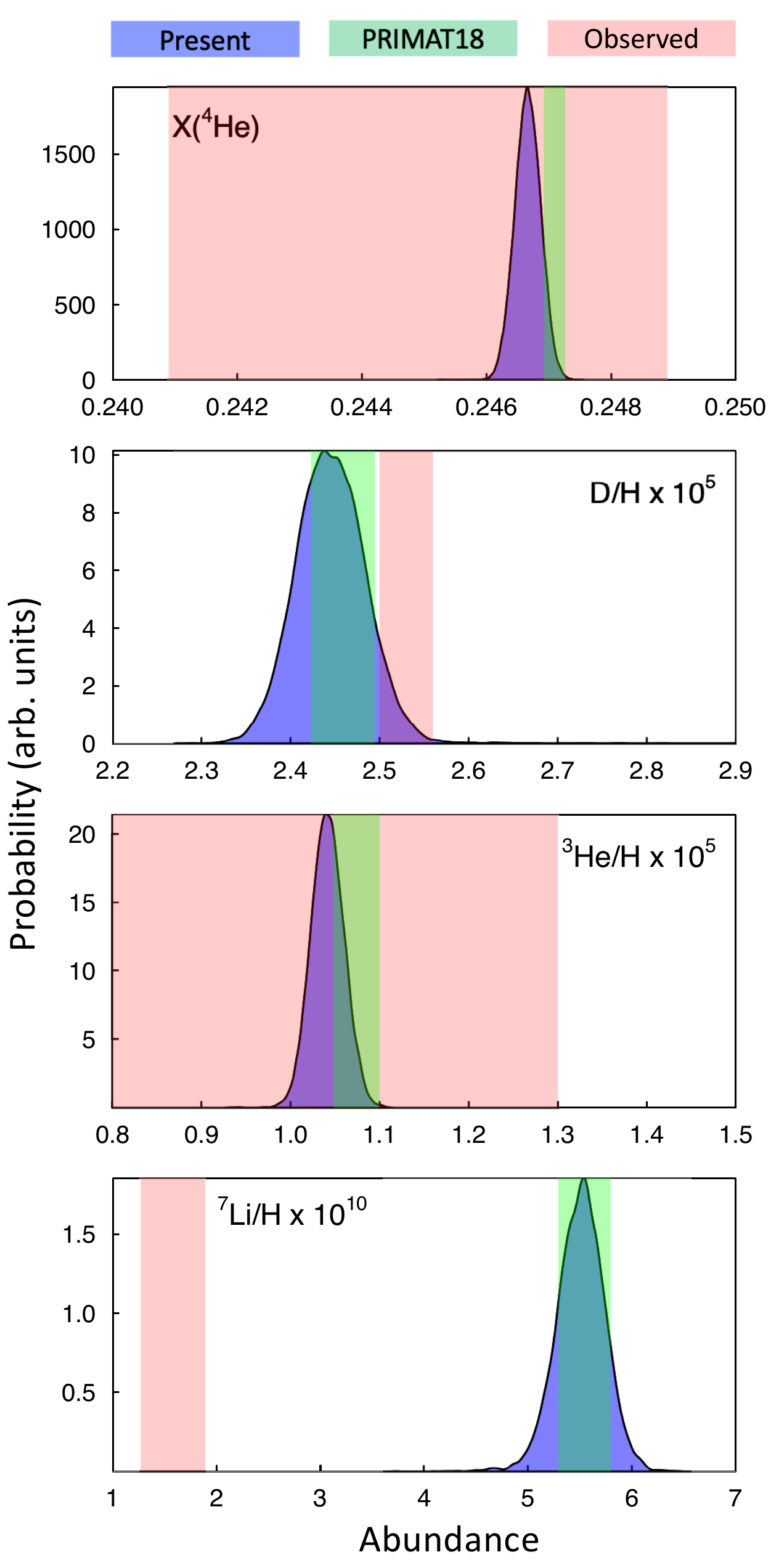}
\caption{Number abundance ratios relative to $^1$H for primordial $^2$H (=D), $^3$He, and $^7$Li. (Blue) Simulated abundance probability densities from the present work. (Green) Simulated abundances reported by \citet{Pitrou2018}. (Red) Observed values (see Table~\ref{tab:abund}). The simulated results are based on 10,000 reaction network samples.}
\label{fig:abundSim}
\end{figure}

Our simulated primordial abundances are listed in Table~\ref{tab:abund} and displayed in Figure~\ref{fig:abundSim}, together with those obtained using the code PRIMAT \citep{Pitrou2018} and the most recently published measured values. 

\begin{deluxetable}{llll}
\tablecaption{Observed and predicted primordial abundances.\tablenotemark{a}}
 \label{tab:abund}
\tablewidth{\linewidth}
\tablehead{ &   Observed\tablenotemark{b}  & Present\tablenotemark{c} & PRIMAT18\tablenotemark{d}  } 
\startdata
Y$_p$                               &  0.2449(40)   &      \tablenotemark{e}  &      0.24709(17) \\ 
(D/H)$_p$ $\times$ $10^{-5}$        &  2.527(30)    &      {\bf 2.445(37)}    &      2.459(36)   \\ 
($^3$He/H)$_p$ $\times$ $10^{-5}$   &  $\leq$1.1(2) &      {\bf 1.041(18)}    &      1.074(26)   \\ 
($^7$Li/H)$_p$ $\times$ $10^{-10}$  &  1.58(31)     &      {\bf 5.52(22)}     &      5.62(25)    \\ 
\enddata
\tablenotetext{a}{For $^4$He, the nucleon fraction (Y$_p$) is given, while for the other species the number abundance ratio relative to $^1$H is listed. Uncertainties are given in parenthesis, e.g., ``0.2449(40)'' stands for ``0.2449$\pm$0.0040.''}
\tablenotetext{b}{From \citet{Aver2015,Cooke2018,Bania:2002wn,sbordone10}.}
\tablenotetext{c}{Recommended values and uncertainties are obtained from the 16th, 50th, and 84th percentiles of the abundance probability densities (shaded blue in Figure~\ref{fig:abundSim}).}
\tablenotetext{d}{From \citet{Pitrou2018}.}
\tablenotetext{e}{No improvement was obtained in the present work compared to the Y$_p$ value of \citet{Pitrou2018}, see text.}
\end{deluxetable}

The $^4$He abundance depends sensitively on the rates of the weak interactions, $n$ $\leftrightarrow$ $p$, for which we adopted the theoretical results of \citet{Pitrou2018} (see Table~\ref{tab:rates}). The width of the abundance distribution, and, thus, the uncertainty of the predicted $^4$He abundance (Table~\ref{tab:abund}), depends mainly on the experimental uncertainty of the neutron lifetime since this value was used to calibrate the theoretical weak interaction rates. Since we adopted the weak interaction rates of \citet{Pitrou2018}, we do not obtain an improved value for the $^4$He nucleon fraction. Therefore, we do not list a present value of Y$_p$ in Table~\ref{tab:abund}.

For $^2$H (top panel in Figure~\ref{fig:abundSim}), present and previous abundance predictions are in agreement. Interestingly, the measured value is larger by about 3\% compared to the simulated ones. This ``tension'' had already been noted by \citet{Coc:2015gt}. The predicted deuterium abundance depends sensitively on the d(p,$\gamma$)$^3$He reaction rate for which we adopt the results of a recent Bayesian analysis (Table~\ref{tab:rates}). New cross section data for this reaction are expected to be published soon by the LUNA collaboration\footnote{The recent experimental cross sections of \citet{Tisma:2019ug} have not yet been used in d(p,$\gamma$)$^3$He rate evaluations. The uncertainties of their four data points are larger than previous results.}. 

For $^3$He (middle panel in Figure~\ref{fig:abundSim}), the simulated abundances agree with the observed upper limit \citep{Bania:2002wn}. Our result is slightly lower than the simulation of \citet{Pitrou2018}, which is explained by our adoption of the most recent $^3$He(d,p)$\alpha$ rate \citep{deSouza:2019gi} compared to their use of the older rate of \citet{Descouvemont2004}.

Finally, the simulated present and previous $^7$Li abundance values (bottom panel in Figure~\ref{fig:abundSim}) are in agreement, despite the fact that both studies employ different $^7$Be(n,p)$^7$Li reaction rates. The recent rate of \citet{de_Souza_2020}, used here, has a significantly larger uncertainty than the rate of \citet{Descouvemont2004} that was used by \citet{Pitrou2018}. Regardless, the simulated $^7$Li abundance exceeds the observed values by a factor of $\approx$ $3$, as has been pointed out before \citep{Cyb03,Coc04,Cuo04}. We will consider correlations next.

\section{Correlations between abundances and reaction rates: Mutual Information} 
\label{sec:correlations}
We mentioned in Section~\ref{sec:mc} that the network Monte Carlo procedure lends itself to studying correlations between simulated abundances and reaction rates. Since final abundances and rate variation factors are stored after each network run, the analysis of correlation scatter plots will reveal the impact of the variation of each rate on the abundance of every nuclide in the network. 
The obvious challenge is to quantify which rates have the largest impact on a specific nuclide abundance. We need to adopt a useful metric to quantify the correlation, compute the metric for $50$ isotopes $\times$ $466$ reactions, and then sort the results according to impact. Pearson's (product-moment correlation coefficient) $r$, which was used by \citet{Coc_2014}, is a measure for the linear correlation between two random variables. \citet{Iliadis:2015gp} suggested to use Spearman's (rank-order correlation coefficient) $r_s$, which quantifies how well the relationship between two variables is described by a monotonic function. However, these metrics are not without problems in the present context, where correlations are frequently neither linear nor monotonic.

Here, we will adopt the {\it mutual information} metric, which originates from information theory \citep{linfoot,10.5555/1146355}. It quantifies how much information is communicated, on average, in one random variable about another when both are sampled simultaneously. For two random variables, $Y$ and $Z$, with values of $\{y_1, y_2, y_3,...\}$ and $\{z_1, z_2, z_3,...\}$, respectively, their mutual information is defined by
\begin{equation}
\label{eq:mi}
MI = \sum_y \sum_z P(y,z) \log \left[ \frac{P(y,z)}{P(y)P(z)} \right]
\end{equation}
where $P(y)$ and $P(z)$ are marginal distributions of $y$ and $z$, respectively, and  $P(y,z)$ is the joint probability density. An important theorem from information theory states that the mutual information between two variables is zero if, and only if, the two random variables are statistically independent. Unlike Pearson's $r$ and Spearman's $r_s$ coefficients, mutual information has no finite upper bound and, therefore, its absolute magnitude has no straightforward interpretation. However, here we are mainly interested in identifying the most important reactions for a given nuclide. For this purpose, the relative magnitude of the mutual information will suffice\footnote{\citet{linfoot} suggested to transform the mutual information by introducing the {\it informational coefficient of correlation}, defined as IC $\equiv$ $\sqrt{1-e^{-2\cdot \textrm{MI}}}$. The coefficient IC lies between $0$ and $1$, equals zero when the two random variables are statistically independent, and equals unity when they are fully correlated. Again, there is no need in the present work to normalize the mutual information, MI.}. 

\subsection{$^4$He abundance}
\label{sec:he4}
\begin{figure}
\includegraphics[width=0.7\linewidth]{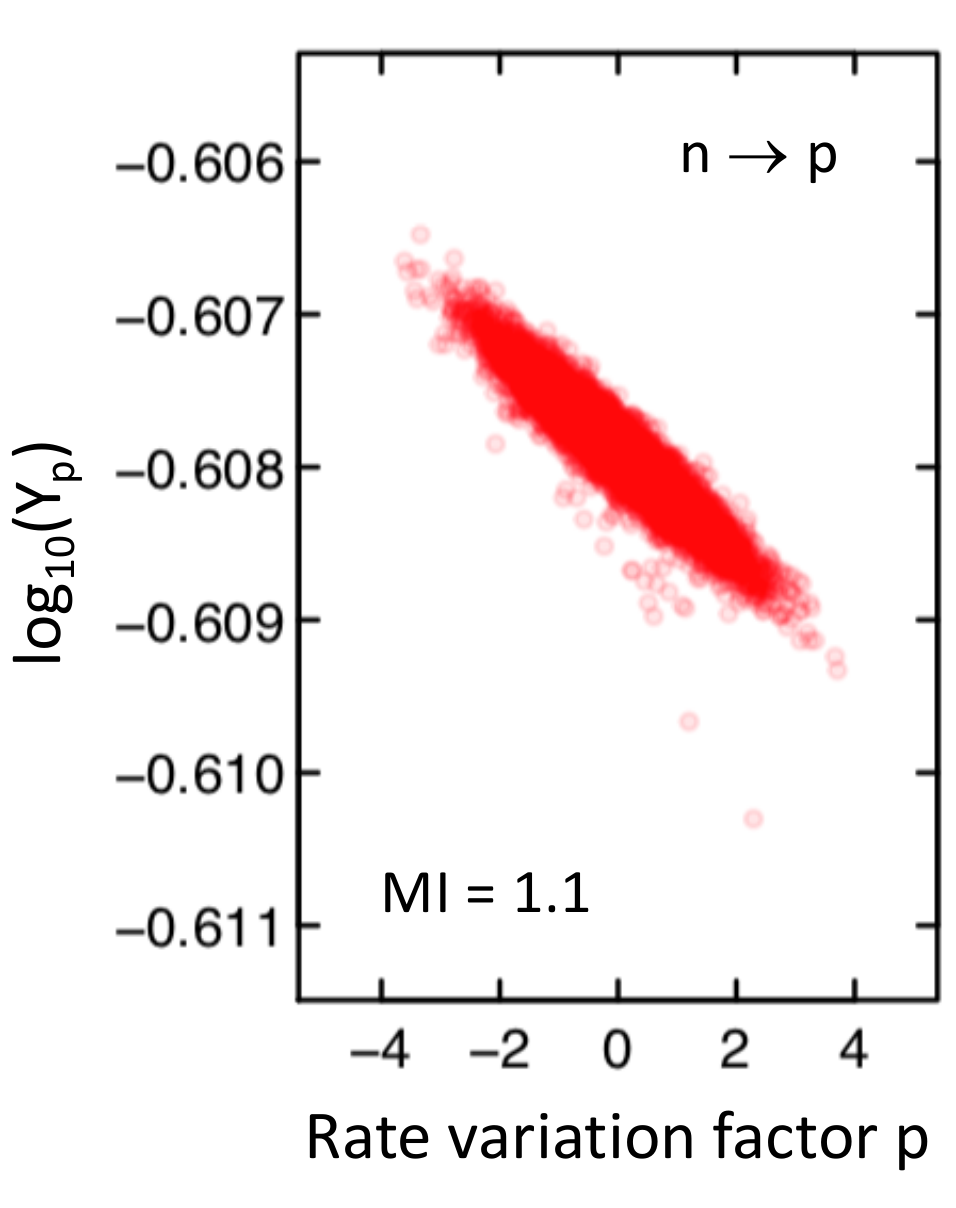}
\caption{Correlation of the simulated primordial $^4$He abundance with the rate variation factor of the weak interactions that transform neutrons into protons (Section~\ref{sec:network}). The value for the mutual information correlation metric is also given (MI $=$ $1.1$). The $^4$He abundance exhibits no noticeable correlation with any other reaction in the network, given the rate uncertainties adopted here. The results are obtained for the same simulations that gave rise to Figure~\ref{fig:abundSim}.}
\label{fig:corr4He}
\end{figure}

The weak interactions, $n$ $\leftrightarrow$ $p$, have the largest impact on the primordial $^4$He abundance (Section~\ref{sec:network}). The correlation between $Y_p$ and the rate variation factor for $n$ $\rightarrow$ $p$ is presented in Figure~\ref{fig:corr4He}. The mutual information amounts to a value of MI $=$ $1.1$. None of the other reactions in the network are noticeably correlated with $Y_p$, i.e., all other mutual information values are MI $\leq$ $0.02$. As already noted in Section~\ref{sec:nuclear}, the weak interaction rate uncertainty is given by the present uncertainty in the neutron decay constant.

Interestingly, if one would assume a factor of $10$ uncertainty for the d(n,$\gamma$)t rate, the $^4$He abundance distribution would display a pronounced tail on the right side of the peak. The published rate uncertainty \citep{Nagai:2006kz} for this reaction is only 8\%, although it is not clear to us how this value was derived. In the present work, we adopt a conservative uncertainty of a factor of $2$. With this assumption, the d(n,$\gamma$)t reaction rate uncertainty has only a negligible impact on the primordial $^4$He abundance. More information about the experiment of \citet{Nagai:2006kz} is given in Appendix~\ref{sec:dng}.

\subsection{$^2$H abundance}
\label{sec:h2}
\begin{figure}
\centering{\includegraphics[width=0.9\columnwidth]{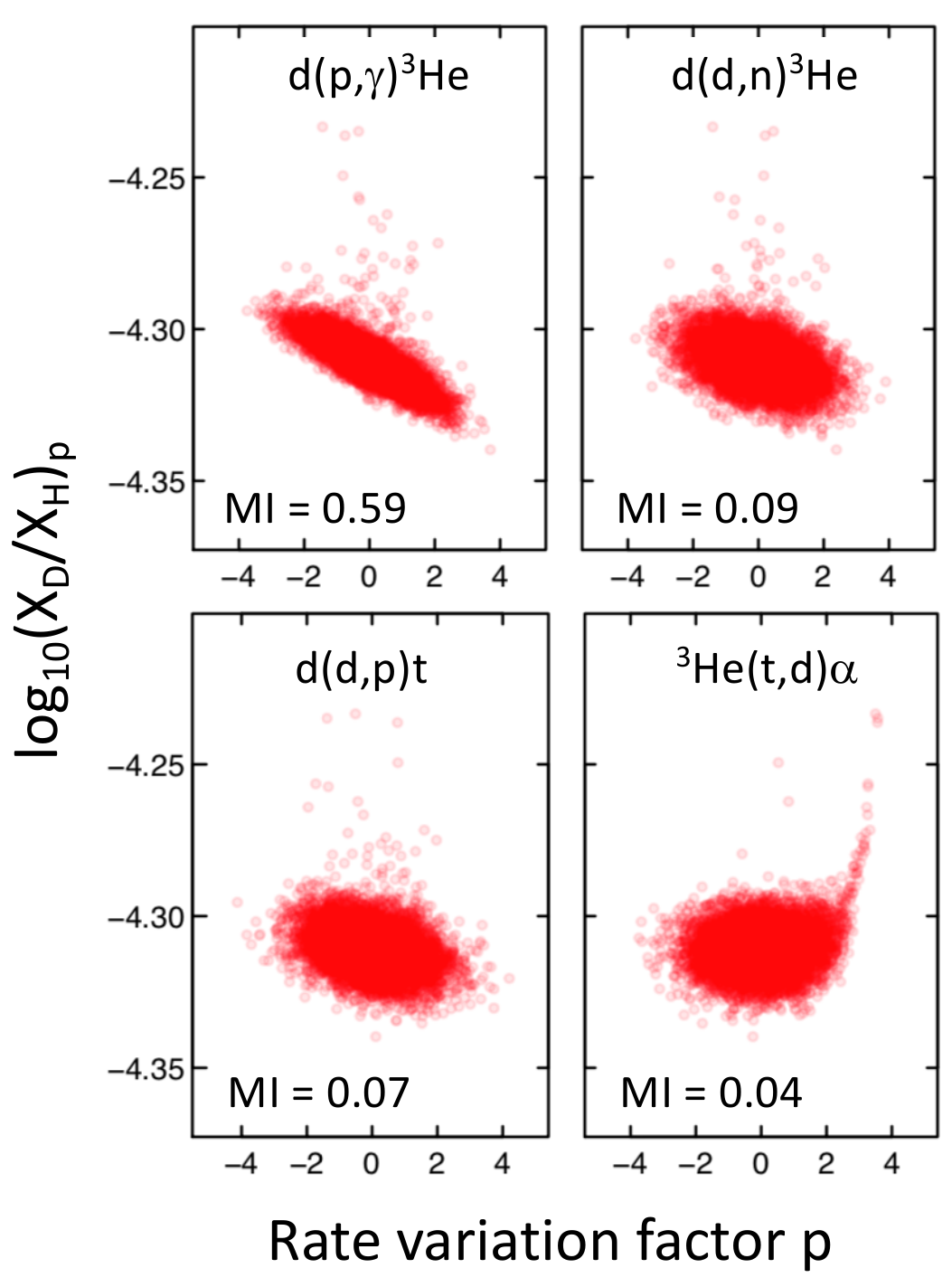}}
\caption{Correlation of the simulated primordial deuterium-to-hydrogen ratio with the rate variation factors of the reactions d(p,$\gamma$)$^3$He, d(d,n)$^3$He, d(d,p)t, and $^3$He(t,d)$\alpha$. The values for the mutual information (MI) are also given. The rate uncertainties of all other reactions in the network have a negligible impact on this abundance ratio. The results are obtained for the same simulations that gave rise to Figure~\ref{fig:abundSim}. The ordinate shows the logarithm of the mass fraction ratio instead of the number abundance ratio.}
\label{fig:corr2H}
\end{figure}

The primordial deuterium abundance is mainly impacted by the rate uncertainties of the d(p,$\gamma$)$^3$He, d(d,n)$^3$He, and d(d,p)t reactions. The correlations are depicted in Figure~\ref{fig:corr2H}. The d(p,$\gamma$)$^3$He reaction is by far the most important among the three processes, as can be seen from its much larger mutual information value (MI $=$ $0.59$) compared to the d(d,n)$^3$He (MI $=$ $0.09$) and d(d,p)t (MI $=$ $0.07$) reactions. The rate uncertainties are listed in Table~\ref{tab:rates} and have been derived using Bayesian techniques for all three reactions. 

As expected, the d(p,$\gamma$)$^3$He rate is negatively correlated with the deuterium abundance: a smaller rate will result in a larger abundance of surviving deuterium nuclei. The present rate uncertainty for the d(p,$\gamma$)$^3$He reaction is $3.7$\%. Decreasing the recommended rate by $9$\% would shift the centroid of the simulated deuterium abundance distribution into the center of the observed range (top panel in Figure~\ref{fig:abundSim}). Clearly, a new and improved measurement of the d(p,$\gamma$)$^3$He reaction is of significant interest (see Section~\ref{sec:network}). 

Figure~\ref{fig:corr2H} also displays the correlation for the $^3$He(t,d)$\alpha$ reaction (MI $=$ $0.04$). The vast majority of network samples reveals an uncorrelated $^2$H abundance, as can be seen from a distribution of points that is symmetric about a rate variation factor of $p$ $=$ $0$. However, a few samples, for large values of $p$, do reveal a correlation. For this reaction, we adopted a rate uncertainty factor of $f.u.$ $=$ $10$. An arbitrary increase of the recommended rate by a factor of $400$ would also shift the peak of the simulated deuterium abundance distribution into the center of the observed range. The status of the rate of this reaction is discussed in Appendix~\ref{sec:he3td}. 

\subsection{$^3$He abundance}
\label{sec:he3}
\begin{figure}
\centering{\includegraphics[width=1.0\columnwidth]{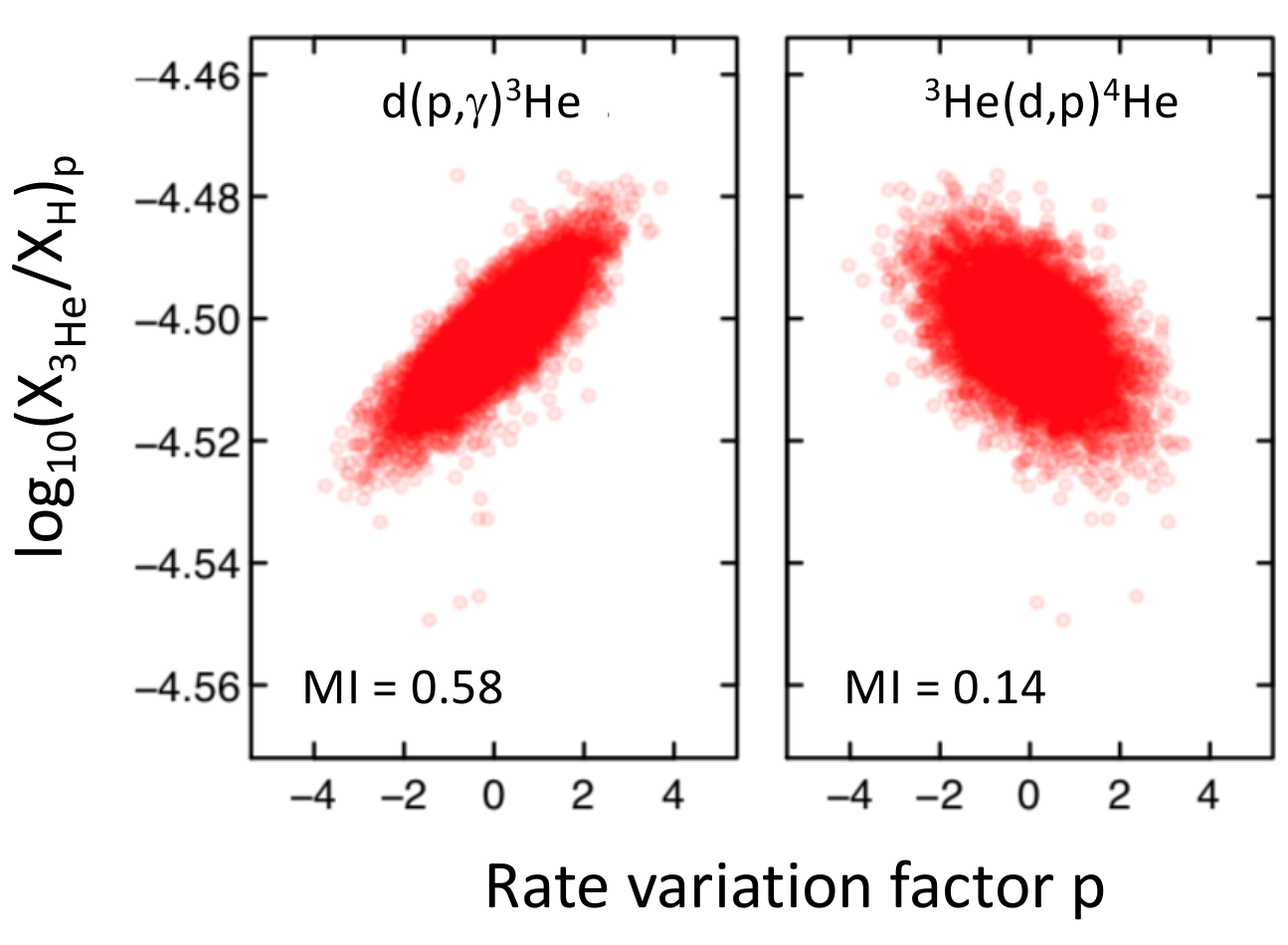}}
\caption{Correlation of the simulated primordial $^3$He-to-hydrogen ratio with the rate variation factors of the reactions d(p,$\gamma$)$^3$He and $^3$He(d,p)$\alpha$. The values for the mutual information (MI) are also given. The rate uncertainties of all other reactions in the network have a negligible impact on the abundance ratio. The results are obtained for the same simulations that gave rise to Figure~\ref{fig:abundSim}. The ordinate shows the logarithm of the mass fraction ratio instead of the number abundance ratio.}
\label{fig:corr3He}
\end{figure}

The primordial $^3$He abundance is mainly influenced by the two reactions that produce and destroy $^3$He, d(p,$\gamma$)$^3$He and $^3$He(d,p)$\alpha$, respectively. Correlations can be found in Figure~\ref{fig:corr3He}. The mutual information values are MI $=$ $0.58$ and MI $=$ $0.14$, respectively, for these two reactions. The reaction rate uncertainties are listed in Table~\ref{tab:rates} and have been derived using Bayesian techniques. None of the other reactions in the network are noticeably correlated with the $^3$He abundance, i.e., all other mutual information values are MI $\leq$ $0.02$.

\subsection{$^7$Li abundance}
\label{sec:be7}
\begin{figure}
\centering{\includegraphics[width=0.9\columnwidth]{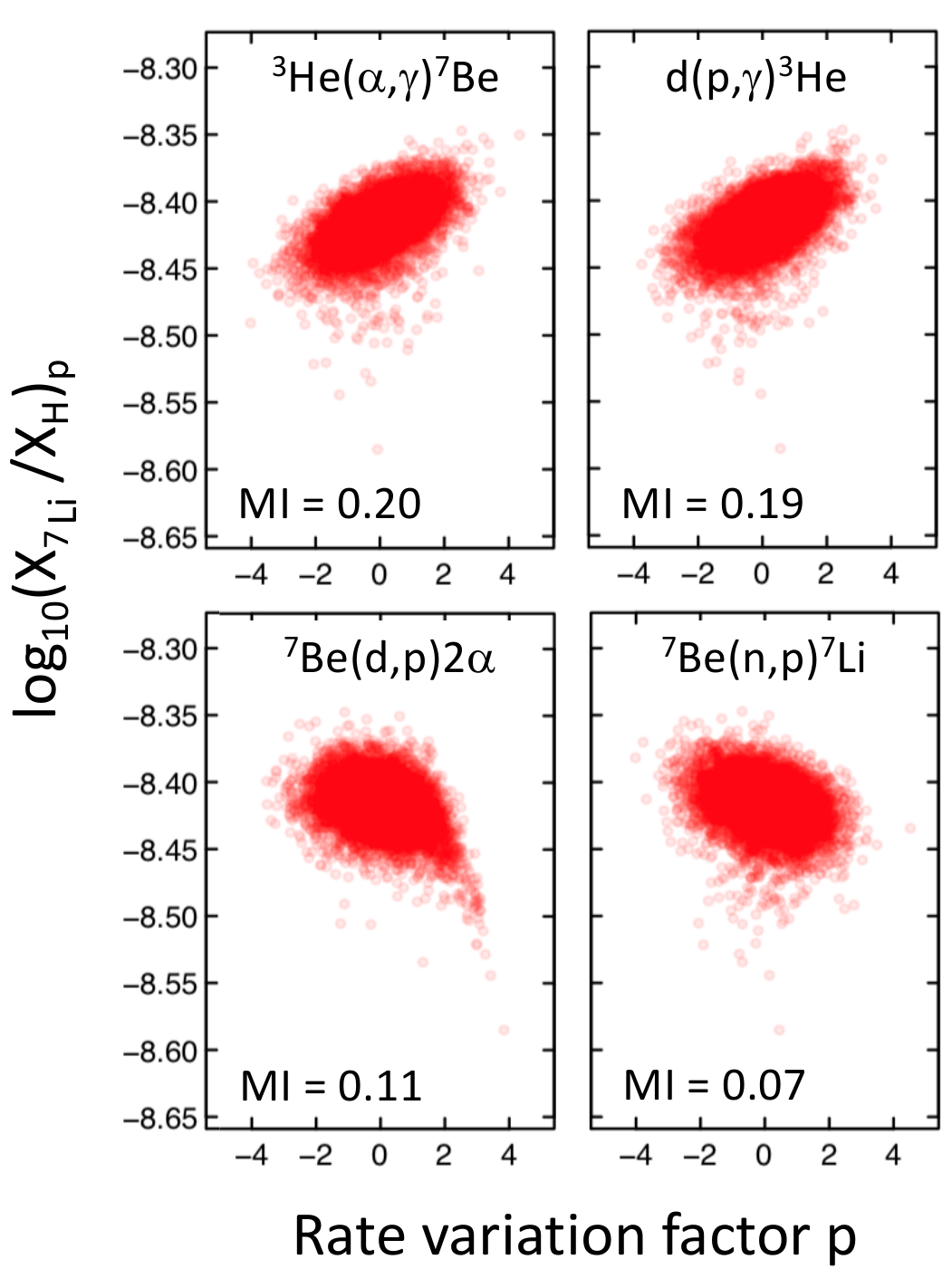}}
\caption{Correlation of the simulated primordial $^7$Li-to-hydrogen ratio with the rate variation factors of the reactions $^3$He($\alpha$,$\gamma$)$^7$Be, d(p,$\gamma$)$^3$He, $^7$Be(d,p)2$\alpha$, and $^7$Be(n,p)$^7$Li. The values for the mutual information (MI) are also given. The rate uncertainties of all other reactions in the network have a negligible impact on the abundance ratio. The results are obtained for the same simulations that gave rise to Figure~\ref{fig:abundSim}. The ordinate shows the logarithm of the mass fraction ratio instead of the number abundance ratio.}
\label{fig:corr7Li}
\end{figure}

Of the primordial lithium synthesized in the early universe, about 95\% originated from radioactive $^7$Be during primordial nucleosynthesis, which later decayed to $^7$Li when the expansion of the universe gave rise to temperatures and densities insufficient to sustain nuclear reactions. The final $^7$Li abundance is mainly impacted by the rate uncertainties of the $^3$He($\alpha$,$\gamma$)$^7$Be, d(p,$\gamma$)$^3$He, $^7$Be(d,p)2$\alpha$, and $^7$Be(n,p)$^7$Li reactions. The correlations are provided in Figure~\ref{fig:corr7Li}. The mutual information values for these four reactions amount to MI $=$ $0.20$, $0.19$, $0.11$, and $0.07$, respectively. The mutual information values of all other reactions in the network are MI $\leq$ $0.03$.

Recently estimated Bayesian rates are available for the $^3$He($\alpha$,$\gamma$)$^7$Be, d(p,$\gamma$)$^3$He, and $^7$Be(n,p)$^7$Li reactions (see Table~\ref{tab:rates}). A more recent $^3$He($\alpha$,$\gamma$)$^7$Be rate has been published by \citet{singh19}, which is based on the calculation of \citet{Dubovichenko18}. However, these results are highly problematic and should not be used in big bang nucleosynthesis simulations, for reasons explained in Appendix~\ref{sec:3heag}. The status of the d(p,$\gamma$)$^3$He reaction rate has already been mentioned in connection with the primordial deuterium abundance (Section~\ref{sec:h2}). The uncertainty of the most recently published $^7$Be(n,p)$^7$Li rate \citep{de_Souza_2020} is 2.1\% at big bang nucleosynthesis temperatures (Table~\ref{tab:rates}).

Figure~\ref{fig:corr7Li} indicates a noticeable tail towards smaller primordial $^7$Li abundances for large values of the $^7$Be(d,p)2$\alpha$ reaction rate variation factor, $p$. In our simulations, we adopted the recently published rate of \citet{Rijal:2019bh}, which has a reported uncertainty of a factor of $\approx$3 at big bang nucleosynthesis temperatures. Increasing the recommended rate by a factor of $\gtrsim200$ would lead to a significant overlap between the predicted and measured abundance values, and, thus, would solve the cosmological lithium problem. Considering the large uncertainties involved in the experiment of \citet{Rijal:2019bh} (30\% in the absolute cross section normalization and $\pm45$~keV in the experimental energy calibration), it is important to remeasure this reaction in the future with improved techniques. More information regarding the present status of the $^7$Be(d,p)2$\alpha$ reaction is provided in Appendix~\ref{sec:be7dp}. 

\section{Unknown systematic bias in reaction rates: Genetic Algorithm} 
\label{sec:bias}
In Section~\ref{sec:mc}, we discussed how to estimate primordial abundances, given a set of thermonuclear rates for the key nuclear reactions (Table~\ref{tab:rates}) that were derived from measured data, by taking into account all known statistical and systematic effects. In agreement with previous work, we find that the simulated primordial lithium abundance exceeds the measured value by a factor of $\approx$3 (Section~\ref{sec:mc}). However, suppose that one or more nuclear reaction rates are subject to some unknown systematic bias, so that their true magnitudes differ significantly from the reported values. Therefore, in this section we ask a different question: {\it By what factors would one need to multiply key reaction rates to reproduce simultaneously all observed primordial abundances?} 

The answer to this question is not simply obtained by independently changing the rates of individual nuclear reactions. We already discussed in Section~\ref{sec:correlations} that, for example, the d(p,$\gamma$)$^3$He reaction impacts the primordial abundances of $^2$H, $^3$He, and $^7$Li (Figures~\ref{fig:corr2H}, \ref{fig:corr3He}, and \ref{fig:corr7Li}, respectively). Therefore, we must consider the simultaneous rate changes of several nuclear reactions.

We will focus on the reactions d(p,$\gamma$)$^3$He, $^3$He($\alpha$,$\gamma$)$^7$Be, $^7$Be(d,p)2$\alpha$, and $^7$Be(n,p)$^7$Li, which correlate most strongly with the primordial $^7$Li abundance (Figure~\ref{fig:corr7Li}). We require that simultaneous rate changes of these reactions reproduce the measured $^2$H, $^3$He, and $^7$Li abundances (Table~\ref{tab:abund}). Although \citet{Bania:2002wn} prefer to report an upper limit for the $^3$He abundance inferred from their observations of distant metal-poor galactic H~II regions, their upper limit value agrees with the simulated primordial $^3$He abundance (Table~\ref{tab:abund}). Therefore, for the discussion in this section, we will follow the suggestion of \citet{Steigman:2007ky} and adopt the upper limit of \citet{Bania:2002wn} as an estimate of the primordial $^3$He abundance.

To investigate this optimization problem, we use a genetic algorithm, which provides a robust means of finding the global extremum in a multi-dimensional parameter space. Genetic algorithms are inspired by the mechanism of biological evolution through natural selection. See, for example, \citet{Goldberg} for details. In brief, the idea is to start with an initial population of unbiased random sets of model parameter values. Each such set represents an individual in the overall population. Depending on their fitness, pairs of individuals are selected for the operations of crossover and mutation to produce offspring until the current generation of individuals has been replaced by a new generation of offsprings. The fitness of each offspring is assessed, and the process repeats itself by evolving through subsequent generations. As a result, the average population fitness increases over time, i.e., a better solution is produced with each generation. The process is terminated when a suitable solution (e.g., a parameter set of predefined fitness) is found or a given number of generations is reached. A major benefit of genetic algorithms is their ability to explore different parts of parameter space simultaneously \citep{Holland}.  

For the genetic algorithm-based optimizer, we used the general-purpose subroutine PIKAIA v1.2\footnote{The public domain source code and accompanying documentation is available at \url{https://www.hao.ucar.edu/modeling/pikaia/pikaia.php}.}, which was first presented in \citet{Charbonneau:1995vn,Gibson:1998vg}. It uses a stochastic rank-based selection scheme, a uniform one-point crossover (occurring with a probability of $0.85$), and a uniform one-point adjustable mutation rate based on fitness. We set the initial, minimum, and maximum mutation probabilities to values of $0.005$, $0.0005$, and $0.25$, respectively. For the reproduction plan, we chose the ``steady-state-delete-worst'' option, where the least fit individual of the parent population  is eliminated and replaced by the offspring.

Fitness is the only point of contact between the genetic algorithm and the problem being solved. We will adopt a figure of merit of the form
\begin{equation}
\chi^2(\boldsymbol{u}) = \sum_{i=1}^3 \frac{[y_{obs,i} - y_i(\boldsymbol{u})]^2}{\sigma_{obs,i}^2}
\label{eq:fitness}
\end{equation}
where $y_i$ and $\sigma_i$ denote the mean values and uncertainties, respectively, of the $^2$H, $^3$He, and $^7$Li measured abundances (Table~\ref{tab:abund}), and $y_i(\boldsymbol{u})$ are the corresponding predictions that depend on the vector of model parameters, $\boldsymbol{u}$. In our context, the model parameters, $u_1$, $u_2$, $u_3$, and $u_4$, are the rate multiplication factors of the reactions d(p,$\gamma$)$^3$He, $^3$He($\alpha$,$\gamma$)$^7$Be, $^7$Be(d,p)2$\alpha$, and $^7$Be(n,p)$^7$Li, respectively. The fitness can then be defined as $f$ $=$ $\chi^{-2}$. If the predictions for all three abundances, (D/H)$_p$, ($^3$He/H)$_p$, and ($^7$Li/H)$_p$, agree with their respective observations within uncertainties, then $f$ $\gtrsim$ $0.33$ according to Equation~(\ref{eq:fitness}). 

\begin{figure}
\centering{\includegraphics[width=1.0\columnwidth]{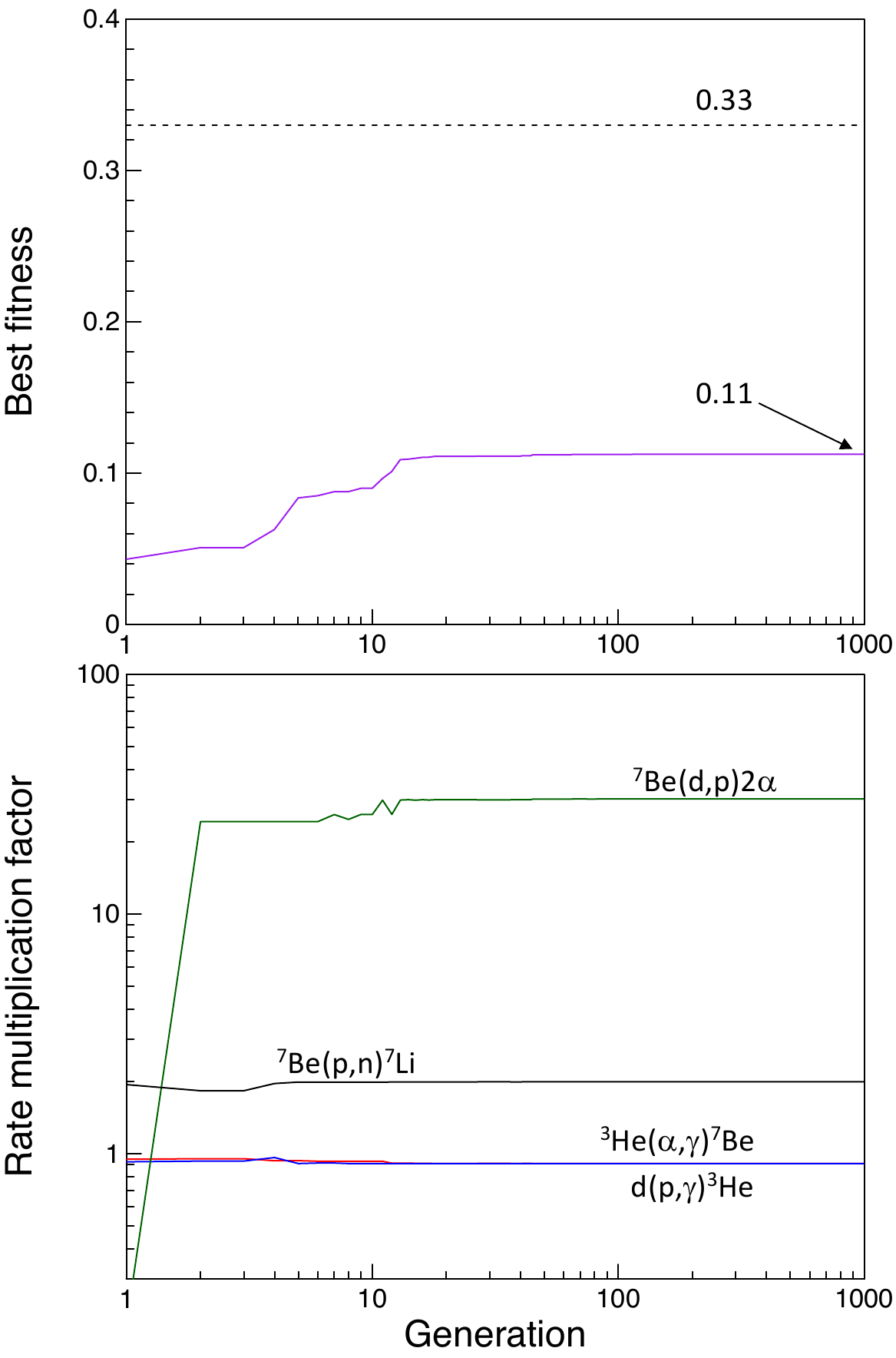}}
\caption{Application of a genetic algorithm to search for possible reaction rate changes that could account for the measured primordial abundances, (D/H)$_p$, ($^3$He/H)$_p$, and ($^7$Li/H)$_p$. The simultaneous rate changes considered are those for the d(p,$\gamma$)$^3$He, $^3$He($\alpha$,$\gamma$)$^7$Be, $^7$Be(d,p)2$\alpha$, and $^7$Be(n,p)$^7$Li reactions. The search ranges for these rates significantly exceed the magnitude of their reported uncertainties (see text). (Top) Evolution of best fitness in each generation. The dashed line ($f$ $=$ $0.33$) indicates the fitness for which the predicted and measured primordial abundances of $^2$H, $^3$He, and $^7$Li would be in agreement. (Bottom) Corresponding evolution of the reaction rate multiplication factors. No acceptable solution ($f$ $\gtrsim$ $0.33$) is found after $1000$ generations. The population size was $100$. 
}
\label{fig:GA}
\end{figure}

First, we restricted the search to the parameter regions $0.9$ $<$ $u_1$ $\leq$ $1.1$, $0.9$ $<$ $u_2$ $\leq$ $1.1$, $0.033$ $<$ $u_3$ $\leq$ $30$, $0.5$ $<$ $u_4$ $\leq$ $2$. These ranges significantly exceed the reported rate uncertainties listed in column 2 of Tables~\ref{tab:rates} and \ref{tab:rates2}. The result of the evolutionary run is shown in Figure~\ref{fig:GA}. The top panel displays the value of $f$ for the fittest individual in each generation. The system is seen to converge after $\approx100$ generations. After $1000$ generations, the maximum fitness value found by the algorithm was only $f$ $=$ $0.11$. In other words, changing the rates of these four reactions within the ranges specified above cannot reproduce the observational abundances. The bottom panel presents the corresponding evolution of the rate multiplication factors. While the factors for d(p,$\gamma$)$^3$He and $^3$He($\alpha$,$\gamma$)$^7$Be are near unity ($u_1$ $\approx$ $u_2$ $\approx$ $1$), those for $^7$Be(d,p)2$\alpha$ and $^7$Be(n,p)$^7$Li are at the upper values of their search ranges ($u_3$ $\approx$ $30$, $u_4$ $\approx$ $2$). Similar results are obtained with different random number seeds.

Since the $^4$He nucleon fraction depends mainly on the weak interaction rates, we have three data points (observations of $^2$H, $^3$He, and $^7$Li) and four parameters (reaction rate multiplication factors). Because our system is underdetermined, we expect, in general, more than one possible combination of parameter values that could reproduce the observations, depending on the constraints (parameter ranges) applied. However, the important point is that the genetic algorithm does not find any acceptable solution at all if the rates of key reactions are varied simultaneously over ranges that are large compared to the reported uncertainties.

Once the search ranges for $^7$Be(d,p)2$\alpha$ and $^7$Be(n,p)$^7$Li ($u_3$ and $u_4$, respectively) are widened even more, the genetic algorithm finds, within only a few generations, many different acceptable solutions ($f$ $\gtrsim$ $0.33$), depending on the magnitude of the search ranges. For example, agreement between observed and simulated abundances is achieved if the $^7$Be(d,p)2$\alpha$ and $^7$Be(n,p)$^7$Li rates are simultaneously multiplied by factors of $u_3$ $\approx$ $0.2$ and $u_4$ $\approx$ $3.4$, respectively, or if the $^7$Be(d,p)2$\alpha$ and $^7$Be(n,p)$^7$Li rates alone are multiplied by factors of $u_3$ $\approx$ $410$ and $u_4$ $\approx$ $4.0$, respectively. Based on the current status of the nuclear physics data, we find such large rate changes to be unlikely. For more information about the $^7$Be(d,p)2$\alpha$ reaction, see Appendix~\ref{sec:be7dp}.

\section{Concluding summary} 
\label{sec:summary}
This work focused on a discussion of thermonuclear reaction rates for big bang nucleosynthesis simulations. We started by summarizing the most recently published reaction rates. Among the twelve nuclear processes that are key to primordial element synthesis, the rates of seven reactions, d(p,$\gamma$)$^3$He, d(d,p)t, d(d,n)$^3$He, $^3$He(d,p)$\alpha$, t(d,n)$\alpha$, $^3$He($\alpha$,$\gamma$)$^7$Be, and $^7$Be(n,p)$^7$Li, have been analyzed using Bayesian techniques (Table~\ref{tab:rates}). Such methods provide results that are less biased when compared to previously applied techniques. Subsequently, we discussed other input to our reaction network, such as the evolution of temperature and density in the early universe, and initial abundances. We then presented simulated primordial abundances, (D/H)$_p$, ($^3$He/H)$_p$, and ($^7$Li/H)$_p$, including their uncertainties. These were obtained using a reaction network Monte Carlo method, i.e., by simultaneously sampling the rates of all reactions in our network according to their rate probability densities. In agreement with previous works, our simulated $^7$Li abundance exceeds the measured primordial value by a factor of $\approx$3. Correlations between predicted abundances and reaction rates were analyzed using the metric of mutual information, which originates from information theory. The most important reactions for the synthesis of $^7$Li are, in decreasing order, according to their mutual information values, d(p,$\gamma$)$^3$He, $^3$He($\alpha$,$\gamma$)$^7$Be, $^7$Be(d,p)2$\alpha$, and $^7$Be(n,p)$^7$Li. These conclusions are based on thermonuclear rates that were derived by taking into account all known statistical and systematic effects in the data analysis. Finally, for these four reactions, we account for {\it unknown} systematic bias in the data by using a genetic algorithm to search for rate multiplication factors that would reconcile all simulated and measured primordial abundances. No such solutions were found within reasonable search ranges, adding weight to previous findings that the solution of the cosmological lithium problem is unlikely to be found within the realm of nuclear physics.

\acknowledgments
We would like to thank Rafael S. de Souza for his help with the genetic algorithm. We would also like to express our gratitude to Robert Janssens for providing valuable feedback. This work was supported in part by NASA under the Astrophysics Theory Program grant 14-ATP14-0007 and by U.S. DOE under contracts DE-FG02-97ER41041 (UNC) and DE-FG02-97ER41033 (TUNL).

\appendix
\twocolumngrid

\section{Comments on specific nuclear reactions}
\label{sec:app}

\subsection{d(n,$\gamma$)t}
\label{sec:dng}
The d(n,$\gamma$)t has been measured by \citet{Nagai:2006kz} in the energy range important for primordial nucleosynthesis. Their three measured cross section data points, at neutron energies of $30.5$~keV, $54.2$~keV, and $521$~keV, have uncertainties between $10$\% and $15$\%, including counting statistics, detector response, correction factors, and absolute normalizations. Their data, including previous results of thermal-neutron capture measurements \citep{PhysRevC.25.2810}, were fit using a theoretical calculation based on the Faddeev approach. However, \citet{Nagai:2006kz} provide insufficient information to assess how this input translates to a reaction rate uncertainty of only $8$\%. For the present simulations, we adopted a more conservative rate factor uncertainty of $f.u.$ $=$ $2$.  

\subsection{$^3$He(t,d)$\alpha$}
\label{sec:he3td}
The $^3$He(t,d)$\alpha$ recommended rate used in the present work was adopted from \citet{CF88}. This rate is presumably the same as the one published in \citet{Fowler:1967tu}, who quote as reference \citet{Youn:1961vk}. The latter work measured the $^3$He $+$ $t$ cross section at triton energies in the range from $150$ to $970$~keV. Their result is in reasonable agreement with \citet{Smith:1963wl}, who measured the cross section at a higher triton energy of $1.9$~MeV, but disagrees with the earlier measurement of \citet{Moak:1953wt}, who obtained a cross section higher by a factor of $\approx$3. \citet{Fowler:1967tu} assigned a rate uncertainty of $\pm30$\% to the $^3$He(t,d)$\alpha$ reaction at all temperatures below $T$ $\leq$ $10$~GK. Although the data are sparse and exhibit significant inconsistencies, it is unlikely that this rate could change by a factor of $\approx400$ required to impact the primordial deuterium abundance significantly, as discussed in Section~\ref{sec:h2}. In the present work, we adopted a conservative uncertainty of a factor $10$ for this rate.

\subsection{The $^3$H\lowercase{e}($\alpha,\gamma)^7$B\lowercase{e} rate}
\label{sec:3heag}
\begin{figure}
\includegraphics[width=.45\textwidth]{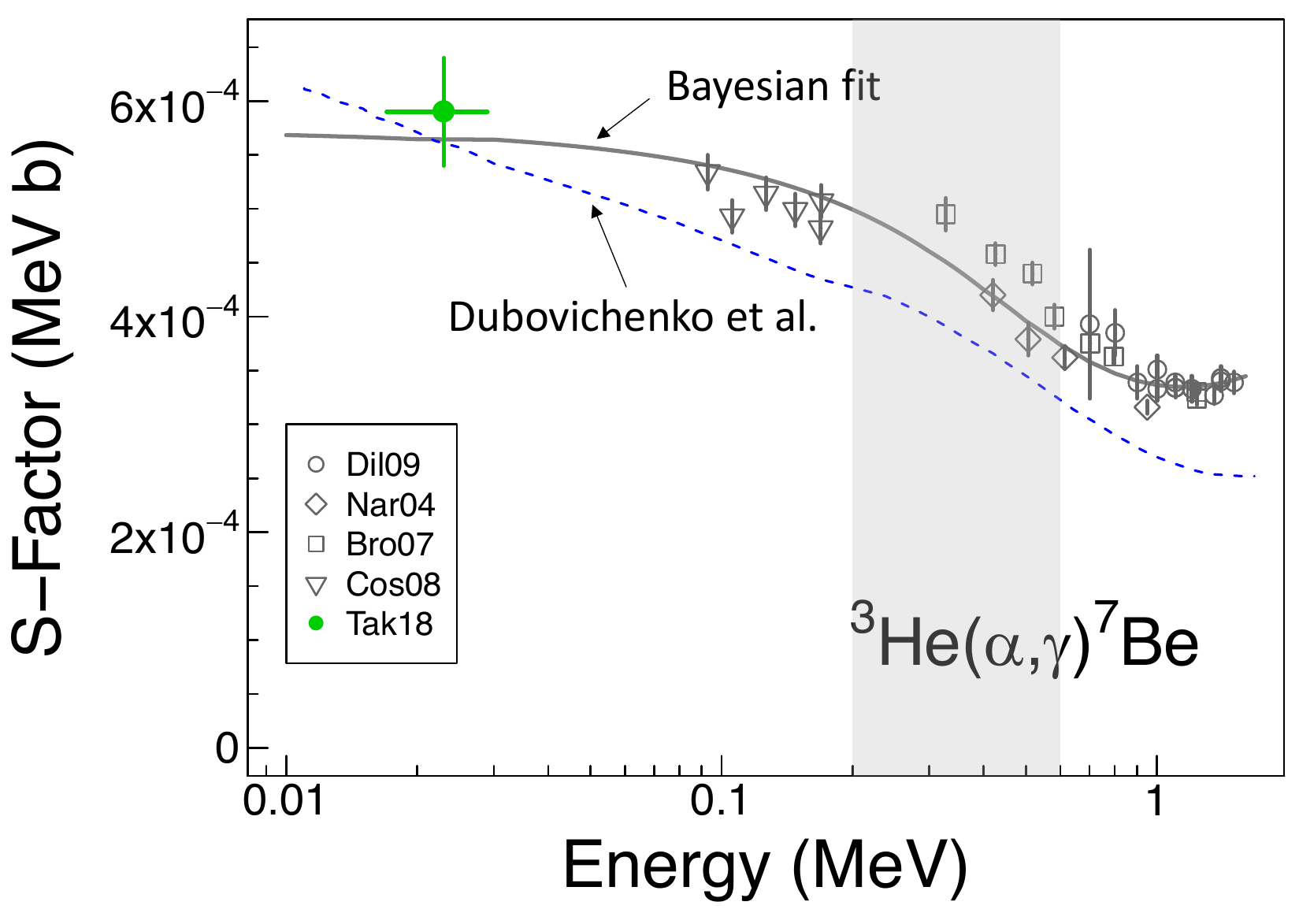} 
\caption{Measured and predicted $^3$He($\alpha$,$\gamma$)$^7$Be S-factors. The modern (published after 2000) data for the direct measurement are shown in black: Dil09 \citep{Dil09}, Bro07 \citep{Bro07}, Nar04 \citep{Nar04}, Cos08 \citep{Costantini:2008it}. The green data point depicts the value obtained indirectly by \citet{TAKACS201878}. The black solid line corresponds to the Bayesian fit of \citet{iliadis16}, which is close to the fits (not shown) of \citet{Cyb08a} and \citet{Ade11}. The dashed blue line indicates the theoretical prediction of \citet{Dub17,Dub19}, which is clearly incompatible with the data shown at energies important for primordial $^7$Be synthesis (shaded region).}
\label{fig:hag1}
\end{figure}

For a long time, measurements of the $^3$He($\alpha$,$\gamma$)$^7$Be reaction have been plagued by systematic effects. Depending on the detection geometry, either the prompt $\gamma$ rays, the $^7$Be decay, or the recoiling $^7$Be nuclei were detected. However, the cross section data from different measurements displayed significant differences. Consequently, the reaction rate based on the available data was rather uncertain \citep{NACRE,Descouvemont2004}. The situation at the time called for new, careful measurements. The S-factors measured subsequently  \citep{Nar04,Bro07,Dil09,Costantini:2008it}, which are displayed in Figure~\ref{fig:hag1} as black data points, are in overall agreement. Also, some groups have applied all three different experimental methods (prompt, decay, and recoil detection), yielding consistent results. Compared to the older measurements (see, for example, Figure~3 of \citet{Nef11}), the modern data are systematically higher and have a significantly smaller dispersion. 

These modern $^3$He($\alpha$,$\gamma$)$^7$Be data were used in a number of studies to derive reaction rates \citep{Cyb08a,Ade11,iliadis16}. The solid black line in Figure~\ref{fig:hag1} displays the fit based on a hierarchical Bayesian model \citep{iliadis16}. It resulted in a reaction rate with an uncertainty of $2.4$\% in the temperature range of big bang nucleosynthesis (Table~\ref{tab:rates}). This rate was adopted in the present work. 

Unfortunately, recent publications \citep{Dub17,Dub19,Singh} have led to confusing claims that cannot be substantiated, as will be explained below. The fit of \citet{iliadis16} did not take into account the single data point of \citet{TAKACS201878}, who inferred the $^3$He($\alpha$,$\gamma$)$^7$Be S-factor at solar energy from the solar neutrino flux. This indirect determination is not on par with the direct measurements mentioned above because it is difficult to assess the systematic uncertainties involved. This data point, shown in green in Figure~\ref{fig:hag1}, reveals large uncertainties and is located far away from the energy range important to primordial nucleosynthesis. However, motivated by this result, a new theoretical S-factor for the $^3$He($\alpha$,$\gamma$)$^7$Be reaction has been proposed by \citet{Dub17,Dub19}, which is shown as the dashed blue line. It can be seen that the fit of \citet{iliadis16} (and those of \citet{Cyb08a,Ade11}) and the theoretical prediction of \citet{Dub17,Dub19} both agree with the (green) data point at solar energy. However, importantly, the theoretical S-factor of \citet{Dub17,Dub19} clearly disagrees with all of the modern data at the energies most relevant for big bang nucleosynthesis. In conclusion, the S-factor of \citet{Dub19} should not be used in primordial nucleosynthesis calculations. It underpredicts the data by about $15$\%, implying a reduction in the simulated primordial lithium abundance by a similar amount.   


The reason why we have discussed the results of \citet{Dub17,Dub19} in some detail is because their S-factor was adopted in the big bang nucleosynthesis simulations of \citet{Singh}. The latter authors reported a smaller predicted lithium abundance, ($^7$Li/H)$_p$ $=$ $(4.447\pm0.067)\times10^{-10}$, which appears to be in lesser disagreement with the observed value (Table~\ref{tab:abund}). However, as explained above, their result is erroneous because it rests on an inappropriate S-factor extrapolation from the solar region to the energy range of primordial nucleosynthesis. 
%

\subsection{The $^7$B\lowercase{e}(\lowercase{d},\lowercase{p})2$\alpha$ rate}
\label{sec:be7dp}
The $^7$Be(d,p)$2\alpha$ reaction was recognized for some time as the most promising one for solving the cosmological lithium problem \citep{Coc04}. Its rate required an increase by about two orders of magnitude compared to the rate of \citet{CF88} to significantly reduce the $^7$Li yield; see Figure~4 in \citet{Coc04} and Table~I in \citet{Coc12}. The \citet{CF88} rate was adopted from \citet{Par72}, who assumed a constant S-factor of 100~MeVb based on the experimental data of \citet{Kav60}. \citet{Par72} writes: {\it ``Experimental measurements of the differential cross-section for this reaction were made by Kavanagh (1960) for deuteron energies from 700 to 1700 keV. Lacking complete angular distributions, these data can be approximately converted to total cross sections by multiplying by 4$\pi$ and (2) by multiplying by a factor of 3 to take into account contributions from higher excited states in $^8$Be.''} Unobserved resonances that could significantly increase the cross section were proposed by \citet{Cha11,Bro12}. Until very recently, this possibility seemed unlikely based on measurements of the average cross section \citep{Ang05} and investigations of the properties of candidate resonances \citep{OMa11,Sch11,Kir11}. Notice that the experiment by \citet{Ang05} was only sensitive to the $^7$Be(d,p)$^8$Be(2$\alpha$) reaction channel. 

In a recent experiment, \citet{Rijal:2019bh,RijalErrata} found a new resonance in the previously undetected $^7$Be(d,$\alpha$)$^5$Li(p$\alpha$) channel. Their experimental S-factor is much higher compared to the results of \citet{Kav60} and \citet{Ang05}. This conclusion was supported by preliminary results of \citet{Ino18}, showing an increase of the S-factor around $E_\mathrm{cm}$ $\approx$ $0.3$~MeV. 

\begin{figure}
\includegraphics[width=.45\textwidth]{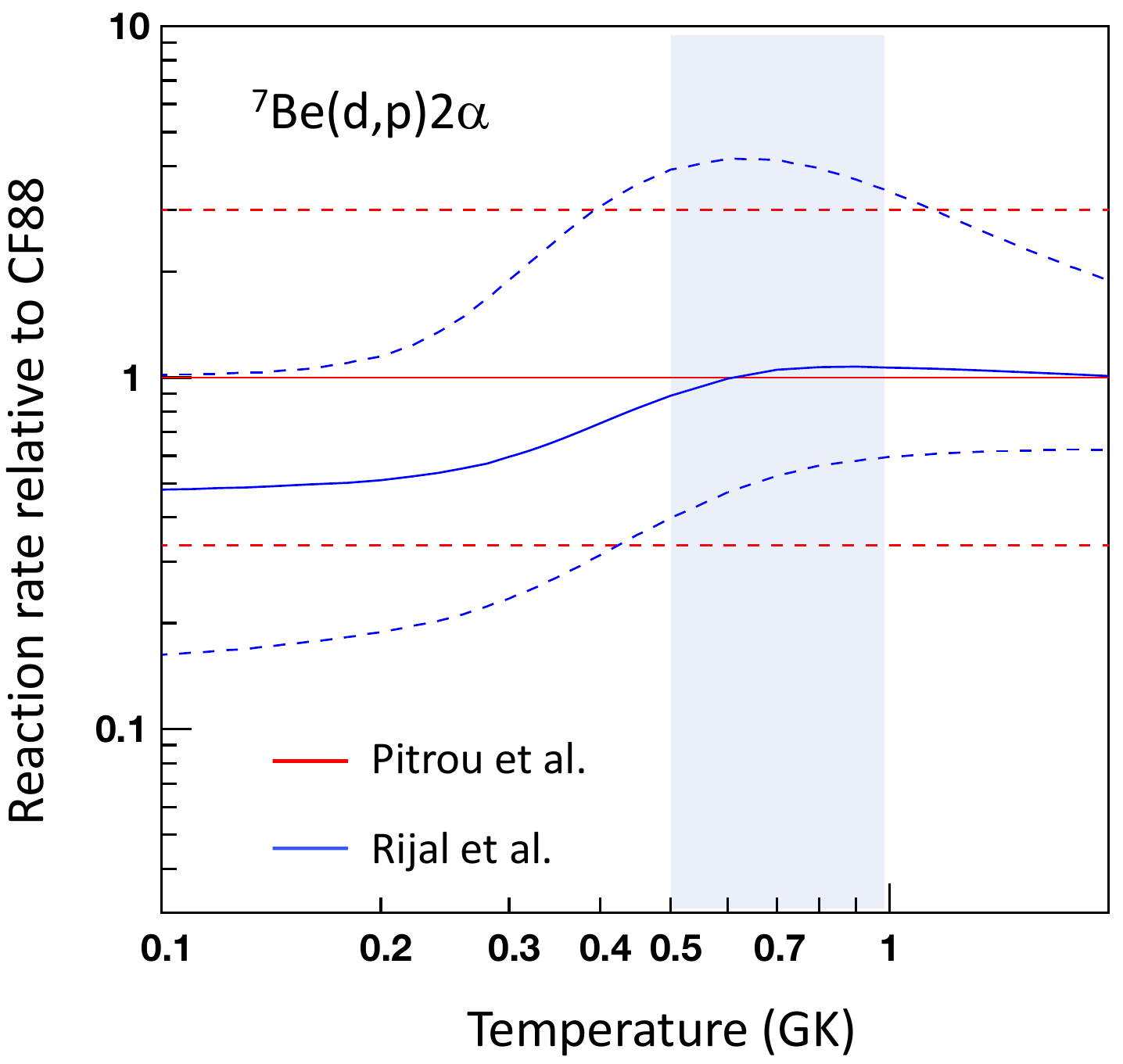} 
\caption{Rates of the $^7$Be(d,p)$2\alpha$ reaction normalized to the recommended rate (``CF88'') of \citet{CF88}. Solid and dashed lines correspond to recommended rates and rate uncertainties, respectively. (Red) From \cite{Pitrou2018}; since no uncertainty estimate was provided by \citet{CF88}, a rate uncertainty factor of $f.u.$ $=$ $3$ was assumed. (Blue) From \cite{Rijal:2019bh}. Notice that the results shown in red and blue are close to each other in the temperature range (shaded band) of primordial $^7$Be synthesis.}
\label{fig:bedprate}
\end{figure}

Although the experiment by \citet{Rijal:2019bh} provided valuable data at temperatures relevant to primordial nucleosynthesis, the magnitude of the reaction rates in the temperature range of primordial nucleosynthesis changed only modestly compared to earlier estimates. \citet{omeg15,Gai19,CD19} pointed out that the recommended rate of \citet{Rijal:2019bh} is close to the result of \citet{Par72,CF88}, who had introduced an approximate rate multiplication factor of $3$. Also, the rate uncertainty reported in \citet{Rijal:2019bh} is close to the factor of $3$ uncertainty adopted by \citet{Pitrou2018}. Figure~\ref{fig:bedprate} displays the rates of \citet{Rijal:2019bh} as blue lines and those adopted by \citet{Pitrou2018} as red lines. The shaded region indicates the relevant temperature range. The rate difference is much smaller than the two orders of magnitude required for a significant change in the predicted primordial lithium abundance; see Table~\ref{tab:abund} and \citet{omeg15,Fields_2020,CD19}. We also note that the claim in \citet{RijalErrata} of a $\approx5$\% reduction in the primordial lithium abundance when the $^7$Be(d,p)$2\alpha$ rate is set equal to zero is most likely erroneous. In our network calculations, the reduction amounts to only 0.75\%.

It is, nevertheless, of interest to remeasure the $^7$Be(d,p)$2\alpha$ reaction and to reduce the uncertainties in resonance energy and cross section, as suggested by \citet{Rijal:2019bh}.

\bibliographystyle{aasjournal}
\bibliography{ref}
\end{document}